\documentclass[twocolumn,aps,prl,showpacs]{revtex4}

\usepackage{epsfig}

\begin{document}

\title {Localized Bosonic Modes in Superconductors}
\author{Dirk K.~Morr and Roy H. Nyberg}
\affiliation{Department of Physics, University of Illinois at
Chicago, Chicago, IL 60607}
\date{\today}
\begin{abstract}

We show that a localized bosonic mode acts as a new type of
``defect" in $s$- and $d_{x^2-y^2}$-wave superconductors. The mode
induces bound or resonance states, whose spectral signature are
peaks in the superconductor's density of states (DOS). We study
the peaks' shape and energy as a function of temperature and the
mode's frequency and lifetime. We identify several characteristic
signatures of the localized mode that qualitatively distinguishes
its effects from those of magnetic or non-magnetic impurities.

\end{abstract}


\maketitle

The study of local defects or impurities in superconductors has
attracted significant experimental \cite{Yaz97,exp} and
theoretical \cite{theo1,theo2,Bal96} interest over the last few
years. These studies have proven particularly important for
further elucidating the nature of the superconducting (SC) pairing
mechanism in unconventional superconductors. In particular, recent
scanning tunnelling microscopy (STM) experiments provided a
detailed picture of the frequency and spatial dependence of defect
induced resonance states in the high-temperature superconductor
(HTSC) Bi$_2$Sr$_2$CaCu$_2$O$_{8+\delta}$ \cite{exp}, and the
triplet superconductor Sr$_2$RuO$_4$ \cite{Davispc}. Several
theoretical scenarios for the physical origin of these impurity
states have been proposed, ranging from electronic scattering off
classical impurities \cite{theo1,Bal96} to the onset of
Kondo-screening \cite{theo2}.

In this Letter we investigate a new type of ``defect", a localized
bosonic mode, and study its effect on the local electronic
structure in $s$- and $d_{x^2-y^2}$-wave superconductors. The
study of localized modes, and in particular {\it phonon modes} in
normal metals, has been of great interest in the context of
molecular spectroscopy \cite{Jak66,review,Sca67} (or inelastic
electron tunnelling). It was recently argued that a localized {\it
magnetic mode} could also arise from unscreened magnetic moments
in a $d_{x^2-y^2}$-wave superconductor \cite{Bal02}. The nature of
the mode is only relevant for the present study to the extent that
it determines the form of its coupling to the electronic degrees
of freedom. While theoretical studies \cite{Sca67,Bal02} have so
far only focused on the Born-limit of weak electronic scattering
off the localized mode, we show that the strong-scattering,
unitary limit gives rise to an abundance of novel phenomena. In
particular, we show that the presence of a localized mode leads to
the appearance of fermionic bound or resonance states whose
spectral signature are peaks in the SC DOS. We study the peaks'
energy and shape as a function of the mode's characteristic
frequency $\omega_0$, lifetime $\Gamma^{-1}$, and temperature. It
is of particular interest that the peaks move to higher energies
with increasing $\omega_0$ or $\Gamma$, but are shifted to lower
energies with increasing temperature, $T$. In addition, the
temperature evolution of the peaks differs {\it qualitatively} if
its frequency is smaller or larger than $\Delta_0 - \omega_0$,
where $\Delta_0$ is the maximum SC gap. Moreover, we show the mode
also induces a ``dip" in the DOS at frequencies
$\pm(\Delta_0+\omega_0)$. This dip, together with the temperature
evolution of the DOS, {\it qualitatively} distinguishes the
effects of a localized mode from those of non-magnetic or static
magnetic impurities \cite{theo1,Bal96}. These characteristic
differences therefore provide an important tool for future
experiments to further study and clarify the nature of ``defects".

%
\begin{figure}[h]
\epsfig{file=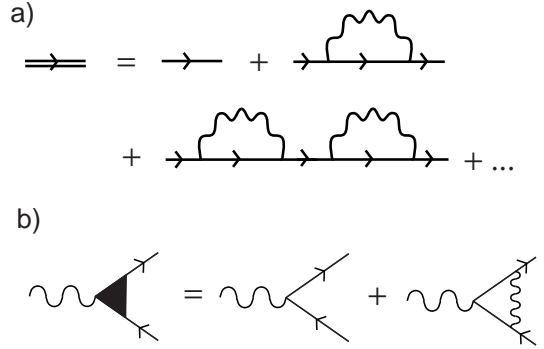,width=7.0cm} \caption{{\it (a)} The full
electronic Greens function. {\it (b)} Lowest order vertex
corrections.} \label{chisigma}
\end{figure}
Starting point for our calculations is the Hamiltonian
\begin{eqnarray}
H&=&\sum_{\bf k} \epsilon_{\bf k} c^{\dagger}_{\bf k,\sigma}
c_{\bf k,\sigma} + \sum_{\bf k} \Delta _{\bf k}c_{{\bf
k},\uparrow}^{\dagger }c_{{\bf k},\downarrow}^{\dagger} + h.c.
\nonumber \\
& & \quad + \omega_0 b^{\dagger}_0 b_0  + \sum_{\bf k,q,\sigma}
g_{\sigma} \left(b^{\dagger}_0 + b_0 \right) c^{\dagger}_{\bf
k,\sigma} c_{\bf q,\sigma}   \ . \label{H1}
\end{eqnarray}
Here, $c_{k}^{\dagger}, b^{\dagger}_0$ are the fermionic and
bosonic creation operators, respectively. We consider a
two-dimensional electronic system with normal state dispersion
$\epsilon_{\bf k}$ and SC gap $\Delta _{\bf k}$. The bosonic mode
with frequency $\omega_0$  is localized at ${\bf r}=0$. We allow
for a spin dependence of the scattering vertex, $g_{\sigma}$, and
for simplicity consider only an on-site interaction. In the
following, we study a single bosonic mode with no internal degrees
of freedom, i.e., spin $S=0$. The extension of our results to $S
\geq 1$, in which case electronic spin-flip scattering is allowed,
is straightforward and will be discussed in a future publication
\cite{Nyb03}. The retarded bosonic Greens function $D_R(\omega)$
in frequency space is given by
\begin{equation}
D_R(\omega)=\frac{\alpha}{(\omega+i\Gamma/2)^2-\omega_0^2} \ .
\label{DR}
\end{equation}
where $\alpha$ depends on $\Gamma$ and $\omega_0$, and is
determined via
\begin{equation}
1+2n_b(\omega_0)=-\int_{-\infty}^{\infty} \frac{d\omega}{\pi} \,
n_b(\omega_0) {\rm Im} \, D_R(\omega) \ .
\end{equation}
We assume that $\omega_0$ and the mode's lifetime, $\Gamma^{-1}$,
are determined intrinsically such as in the case of phonon modes
\cite{Jak66}. Possible feedback effects on $\omega_0$ and $\Gamma$
due to the fermion-boson coupling are accounted for (to the extent
that they do not induce a frequency dependence of $\Gamma$) by
considering $\omega_0$ and $\Gamma$ to be effective temperature
independent input parameters of our theory. The unperturbed
(clean) fermionic Greens function is given by $\hat{G}^{-1}_0({\bf
k},i\omega_n)=\left[ i\omega_n \tau_0 - \epsilon_{\bf k} \tau_3
\right] \sigma_0 + \Delta_{\bf k} \tau_2 \sigma_2 \ , $ where
$\sigma_i$ and $\tau_i$ are the Pauli-matrices in spin and
Nambu-space, respectively. Changes in the fermionic Greens
function due to the scattering of the localized mode are accounted
for by summing the infinite series of diagrams shown in
Fig.~\ref{chisigma}a which yields
\begin{eqnarray}
\hat{G}(r,r',\omega_n)&=&\hat{G}_0(r,r',\omega_n) \nonumber \\
& & \hspace{-1cm}+\hat{G}_0(r,0,\omega_n) \hat{S}(\omega_n)
\hat{G}_0(0,r',\omega_n) \ . \label{fullG}
\end{eqnarray}
Here
\begin{equation}
\hat{S}(\omega_n)=\hat{\Sigma}(\omega_n)
\left[\hat{1}-\hat{G}_0(0,0,\omega_n)\hat{\Sigma}(\omega_n)
\right]^{-1} \label{Smatrix}
\end{equation}
and $\hat{\Sigma}$ is the local fermionic self-energy given by
\begin{equation}
\hat{\Sigma}(\omega_n)= \ T \sum_{m} \tau_3 {\hat g} \  {\hat
G}_{0}(0,0,\omega_{n}-\nu_{m})D(\nu_{m}) \ \tau_3 {\hat g} \ .
\label{sigma}
\end{equation}
with ${\hat g}$ being determined by the spin dependence of
$g_{\alpha}$. We show later that vertex corrections (see
Fig.~\ref{chisigma}b) are negligible for the range of parameters
considered here. Based on earlier studies \cite{Bal96,Bal02}, we
expect that the inclusion of non-crossing diagrams in the
calculation of $\hat{G}$ does not qualitatively change our results
and therefore reserve its discussion for a future publication
\cite{Nyb03}. The DOS, $N=A_{11}+A_{22}$ with $A_{ii}({\bf
r},\omega)=-2{\rm Im}\, \hat{G}_{ii}({\bf r},\omega+i\delta)$ is
obtained numerically from Eqs.(\ref{fullG})-(\ref{sigma}) with
$\delta=0.2$ meV.

\begin{figure}[t]
\epsfig{file=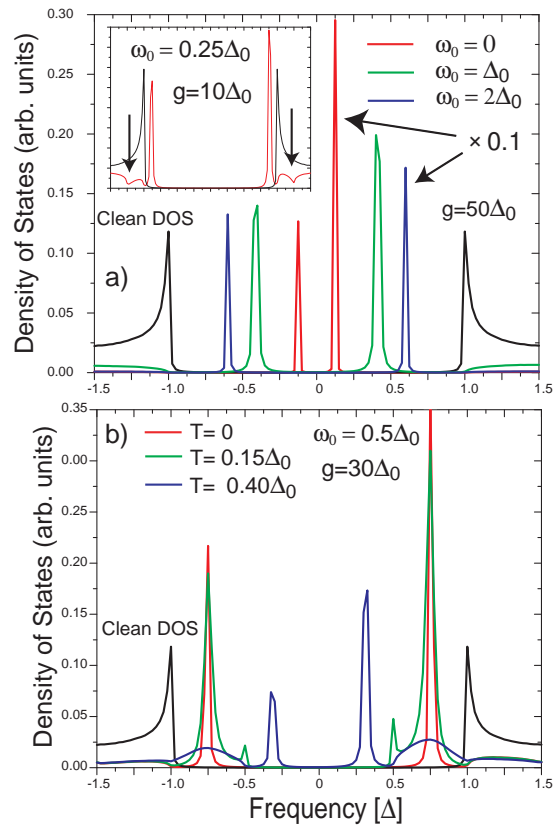,width=7.5cm} \caption{DOS at the site of the
localized mode (${\bf r}=0$) with $\Gamma=0^+$. {\it (a)} $T=0$
and several values of $\omega_0$. The curves for $\omega_0=0$ and
$\omega_0=2\Delta_0$ are multiplied by $0.1$. Inset: arrows
indicate the ``dip" in the DOS. {\it (b)} $\omega_0=0.5\Delta_0$
and several $T$. } \label{swave1}
\end{figure}

As a frame of reference, we first consider a localized mode in an
$s$-wave superconductor with normal state dispersion
$\epsilon_k=k^2/2m - \mu$, $m^{-1}/\Delta_0=15$, Fermi momentum
$k_F=\pi/2$, and a momentum independent SC gap $\Delta_0$. For a
spin-independent coupling, ${\hat g}=g {\hat 1}$, no bound states
are induced, similar to the case of a non-magnetic impurity in an
$s$-wave superconductor \cite{Shiba68}. In contrast, if the
coupling is spin-dependent the mode induces a bound state, with
poles in the $\hat{S}$-matrix leading to two peaks in the DOS. For
simplicity we take $g_{\uparrow}=g$, and $g_{\downarrow}=-g$,
i.e., ${\hat g}=g \sigma_3$, and in Fig.~\ref{swave1} present the
DOS at ${\bf r}=0$ for $\Gamma=0^+$ as a function of $\omega_0$
and temperature.
%
%
For $T=0, \omega_0 = 0$ (see Fig.~\ref{swave1}a) the frequencies
$\pm \Omega_b$ of the bound state peaks are given by
$\Omega_b/\Delta_0= D_-/D_+$ where $D_\pm=1 \pm (gm/4)^2$. This
result is similar to the energy of a bound state induced by a
static magnetic impurity \cite{Shiba68}. Note that increasing $g$,
i.e., stronger scattering, leads to a decrease of $\Omega_b$. For
$\omega_0, T \not = 0$ the bound state frequencies are shifted
from $\Omega_b$, and in the limit $T \ll \Delta_0$ are determined
by a transcendent equation. It is too cumbersome to be presented
here, but can be simplified in the limit $\Omega_b, \omega_0 \ll
\Delta_0$, where we find to leading order in $\omega_0/\Delta_0$
and $n_B(\omega_0)$
\begin{equation}
\Omega_b(T,\omega_0)=\Omega_b + \omega_0/\pi
-n_B(\omega_0)\Delta_0 \ . \label{omb2}
\end{equation}
Note, that the localized mode effectively behaves as an
``oscillating vector" with Ising symmetry (for a single bosonic
degree of freedom) and frequency $\omega_0$. When $\omega_0$
increases (and in particular, when it becomes comparable to the
rate of electronic scattering, $\mu/\hbar$) subsequent scattering
events begin to cancel each other; the effective scattering
becomes weaker, and the peaks are shifted to higher energies. In
particular, for $\omega_0 \rightarrow \infty$ the bound state
vanishes and the DOS becomes that of the unperturbed systems
\cite{Nyb03}. This picture provides an explanation for our
analytical results in Eq.(\ref{omb2}) and fully agrees with the
numerically obtained DOS shown in Fig.~\ref{swave1}a for $T=0$ and
several values of $\omega_0$. A non-zero $\omega_0$ also leads to
a ``dip" in the DOS at $|\Omega|=\Delta_0+\omega_0$, as shown in
the inset of Fig.~\ref{swave1}a. This dip is a characteristic
signature of the localized mode, which distinguishes it from a
static magnetic impurity. The origin of the dip lies in the poles
of the retarded self-energy, $\hat{\Sigma}_R(k,\Omega)$ at
$|\Omega|=E_k+\omega_0$.  The imaginary part of the retarded local
self-energy, $\hat{\Sigma}_{loc}$, then possesses a threshold
energy, $\Omega^+_c=\Delta_0+\omega_0$, with Im$\hat{\Sigma}_{loc}
\sim 1/\sqrt{ (|\Omega|-\omega_0)^2-\Delta_0^2 }$ for $|\Omega|
\geq \Omega^+_c$, and Im$\hat{\Sigma}_{loc} \equiv 0$ for
$|\Omega|<\Omega^+_c$. As a result, Re$\hat{\Sigma}_{loc}$
exhibits a square-root singularity for $|\Omega| \rightarrow
\Omega^+_c -0^+$, which in the denominator of $\hat{S}$ leads to a
suppression, i.e., dip, in the DOS.  This dip is therefore similar
in nature to that observed in angle-resolved photoemission
experiments in the HTSC \cite{arpes}. Finally, we find that the
amplitude of the bound state peaks exhibits spatial oscillations
described by $\cos(2k_Fr) e^{-r/\xi}/r$, where $\xi=k_F/m
\sqrt{\Delta_0^2-\Omega_b^2}$ \cite{Nyb03}.

For non-zero temperatures the $\omega=+\omega_0$ branch of the
bosonic mode becomes populated, opening a new scattering channel
and yielding significant changes in the DOS, as shown in
Fig.~\ref{swave1}b. We find that the specific temperature
evolution of the bound state peaks depends on whether $\Omega_b$
at $T=0$ is smaller or larger than $\Omega^-_c=\Delta_0-\omega_0$.
This distinction arises since $\hat{\Sigma}_R(k,\Omega)$ now not
only possesses poles at $|\Omega|=E_k+\omega_0$, with weight
$1+n_B(\omega_0)$ (for $T, \omega_0 \ll \Delta_0$), but also poles
at $|\Omega|=E_k-\omega_0$ with weight $n_B(\omega_0)$. Hence,
Im$\hat{\Sigma}_{loc} \not = 0$ already for $ |\Omega| \geq
\Omega^-_c$, in contrast to the case at $T=0$. Thus, a bound state
peak with $\Omega^-_c<\Omega_b<\Delta_0$ at $T=0$ becomes damped
with increasing temperature, as shown in Fig.~\ref{swave1}b for
$T=0.15\Delta_0$. At the same time, a new peak appears in the DOS
at $|\Omega|=\Omega^-_c$, since Im$\hat{\Sigma}_{loc}$ now also
possesses square root singularities at $\pm\Omega^-_c$. With
increasing temperature the damping of the bound state peak grows
($T=0.4 \Delta_0$) and its amplitude consequently decreases, while
the newly emerged peak moves towards lower energies. In contrast,
if $\Omega_b \leq \Omega^-_c$ (not shown) the bound state peak
simply moves towards lower energies with increasing temperature,
in agreement with our analytical results in Eq.(\ref{omb2}). This
downshift is expected since the population of the mode's branches
grows with increasing temperature, thus leading to stronger
scattering. Thus, varying $T$ or $\omega_0$ has an opposite effect
on the energies of the bound state peaks in the DOS.

%
%
\begin{figure}[t]
\epsfig{file=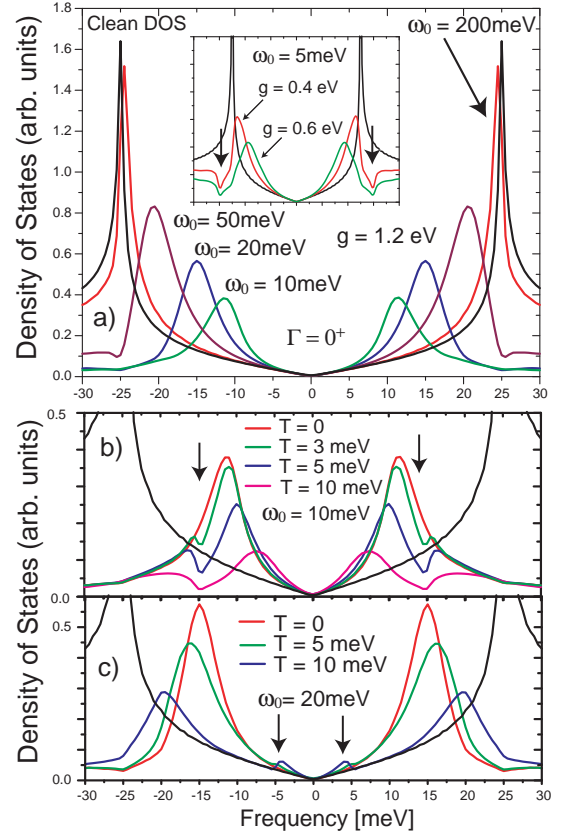,width=7.5cm} \caption{DOS at ${\bf r}=0$ for
$\Gamma=0^+$ and $g=1.2$ eV. {\it (a)} $T=0$ and several values of
$\omega_0$. Inset: arrows indicate the ``dip" in the DOS. {\it
(b)} $\omega_0=10$ meV and several $T$. {\it (c)} $\omega_0=20$
meV and several $T$.} \label{dwave1}
\end{figure}
We next consider a localized mode in a $d_{x^2-y^2}-$wave
superconductor with SC gap $\Delta_{\bf k}=\Delta_0(\cos k_x -\cos
k_y)/2$ and $\Delta_0=25$ meV. To simplify the discussion we
consider a particle-hole symmetric dispersion, $\epsilon_k=-2t
(\cos k_x +\cos ky)$, with $t=300$ meV. A band dispersion with
strong particle-hole asymmetry, characteristic of the HTSC, yields
qualitatively similar results \cite{Nyb03}. In contrast to the
$s$-wave superconductor, a localized mode induces a resonance
state even with a spin-independent coupling, $g_{\alpha}=g$. These
resonances are a hallmark of the strong coupling theory, and are
not observed in the weak scattering limit \cite{Bal02}. In
Fig.~\ref{dwave1} we present the DOS at ${\bf r}=0$ as a function
of $\omega_0$ and temperature ($\Gamma=0^+$). For $T,\omega_0 = 0$
(see Fig.~\ref{dwave1}a), the energy of the resonance peaks,
$\Omega_r$, is given by the solution of $\Omega_r= \pm 2 \pi
\Delta_0 / \left\{ g \log \left[|\Omega_r|/(4 \Delta_0)
\right]\right\}$, similar to the energies of resonance states
induced by a non-magnetic impurity \cite{Bal96} (this result was
also independently obtained by Si \cite{QSipc}). For $|\Omega|
\geq \omega_0\not = 0$,
\begin{equation}
{\rm Im}\Sigma_{loc}(\Omega)=-\frac{1}{\pi}
\frac{|\Omega|-\omega_0}{\Delta_0}
K\left(\frac{|\Omega|-\omega_0}{\Delta_0} \right)
\end{equation}
where $K$ is the complete elliptic integral of the first kind,
while Im$\Sigma_{loc} \equiv 0$ for $|\Omega| < \omega_0$. Thus
there exist a real gap in Im$\Sigma_{loc}$, in contrast to
Im$G_{loc}$. Performing a Kramers-Kronig transform, we find for
$\Omega/\omega_0 \ll 1$, Re$\Sigma_{loc} \approx -A \Omega
/\omega_0$ where $A=- \frac{2}{\pi}\int_1^\infty dx \, {\rm
Im}\Sigma_{loc}(x)/x^2$. In the limit $\Omega_r \ll \Delta_0$, the
resulting equation for $\Omega_r$ is
\begin{equation}
\Omega^2_r=\frac{4 \pi \Delta_0 \Omega}{g^2} \left[ A \log\left(
\frac{4 \Delta_0}{|\Omega_r|} \right) \right]^{-1}
\end{equation}
As expected, increasing $\omega_0$ leads to a shift of $\Omega_r$
to higher energies, in agreement with our numerical results shown
in Fig.~\ref{dwave1}a. Moreover, for $\omega_0 \rightarrow \infty$
we again recover the unperturbed DOS (see, e.g., $\omega_0=200$
meV). The logarithmic divergence in $\Sigma_{loc}$ at
$|\Omega|=\Delta_0+\omega_0$ leads to a dip in the DOS (see arrows
in the inset of Fig.~\ref{dwave1}a). A much weaker dip emerges in
the weak scattering limit \cite{Bal02}.

The specific temperature evolution of the resonance peaks again
depends on the relative order of $|\Omega_r|$ at $T=0$ and
$\Omega^-_c=\Delta-\omega_0$, similar to the $s$-wave case. For
$|\Omega_r|<\Omega^-_c$ (see Fig.~\ref{dwave1}b) the resonance
peak is shifted to lower energies with increasing temperature. In
addition, the opening of a second scattering channel for $T \not =
0$ leads to a logarithmic divergence in $\Sigma_{loc}$, and
consequently a dip in the DOS at $|\Omega|=\Omega^-_c=15$ meV, as
indicated by the arrows. In contrast, for
$\Omega^-_c<|\Omega_r|<\Delta_0$ (see Fig.~\ref{dwave1}c), the
same opening of a second scattering channel leads to an increase
in Im$\Sigma_{loc}$ for $|\Omega|>\omega_0$, which in turn shifts
the peaks to higher energies and suppresses their amplitude. The
smaller peaks at $|\Omega|=\Omega^-_c=5$ meV (see arrows) arise
from the opening of the second scattering channel, and the
resulting logarithmic divergence of Im$\Sigma_{loc}$ at
$\Omega^-_c$. Finally, we find that due to the momentum dependence
of the SC gap, the spatial dependence of the resonance peaks (not
shown) \cite{Nyb03} is qualitatively similar to that observed
experimentally near impurities in the HTSC \cite{exp}.

\begin{figure}[t]
\epsfig{file=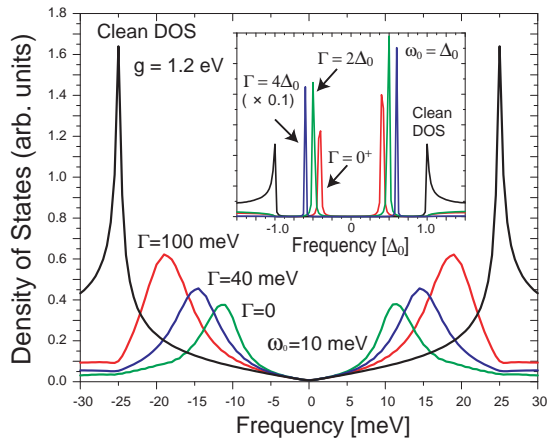,width=7.5cm} \caption{DOS at ${\bf r}=0$ for
$T=0$ and several values of $\Gamma$ for a $d_{x^2-y^2}$-wave and
$s$-wave superconductor (see inset, the curve for
$\Gamma=4\Delta_0$ has been multiplied by $0.1$). } \label{gamma}
\end{figure}

To study the effects of a mode with finite lifetime, $\Gamma \not
= 0$, we present in Fig.~\ref{gamma}, the DOS in a
$d_{x^2-y^2}$-wave and $s$-wave (see inset) superconductor at
$T=0$ and for several values of $\Gamma$.
%
%
In both superconductors the peaks shift to higher energies with
increasing $\Gamma$. Note that a non-zero $\Gamma$ does not {\it
directly} change the lifetime of the induced state, i.e., the
width of the peaks. This is particularly evident for the $s$-wave
case (see inset). A non-zero $\Gamma$ lowers the threshold
frequency for Im$\hat{\Sigma}_{loc} \not = 0$ from
$\Omega^+_c=\Delta_0+\omega_0$ (for $\Gamma=0^+$) to $\Omega_c =
\Delta_0$. However, for $|\Omega| < \Delta_0$, we still have
Im$\hat{\Sigma}_{loc} = 0$ and the width of the bound or resonance
peaks remains unaffected by $\Gamma \not = 0$.

Finally, we consider the lowest order vertex correction, shown in
Fig.~\ref{chisigma}b. For a quadratic normal state dispersion, the
general form of this correction is $\delta g/g=-\left(
\frac{gm}{4} \right)^2 F(\Omega_m,\nu_n,\omega_0,\Delta_0)$ where
$\Omega_m, \nu_n$ are the bosonic and fermionic Matsubara
frequencies, respectively, and $F$ is a bounded function of
$O(1)$. In particular, for an $s$-wave superconductor, we have
$F=+1$ for $\Omega_m,\nu_n,\omega_0 \gg \Delta_0$, and $F=-1$ for
$\Omega_m,\nu_n,\omega_0 \ll \Delta_0$, implying that the bare
vertex is enhanced (suppressed) for frequencies smaller (larger)
than the SC gap. In contrast, in a $d_{x^2-y^2}$-wave
superconductor we obtain $F \rightarrow 0$ for
$\Omega_m,\nu_n,\omega_0 \ll \Delta_0$, and vertex corrections
become irrelevant. Thus, vertex corrections can in general be
neglected for $gm/4 < 1$, which applies to all cases considered
above.

In conclusion, we consider the effects of a localized bosonic mode
on the electronic structure in $s$- and $d_{x^2-y^2}$-wave
superconductors. The mode acts as a new type of ``defect", leading
to the emergence of bound or resonance states. We identify several
characteristic features, such as a ``dip" in the DOS and the
temperature dependence of the mode induced peaks, that
qualitatively distinguishes the effects of a localized mode from
those of static impurities. This result provides further insight
into the nature of impurities observed in STM experiments.

We would like to thank A.~Abanov, A.~Balatsky, J.C.~Davis, A.~de
Lozanne and Q.~Si for stimulating discussions.

\end{document}